%% file: main.tex
\documentclass{article}
\PassOptionsToPackage{numbers,compress}{natbib}
\usepackage[preprint]{neurips_2026}

\usepackage[utf8]{inputenc}
\usepackage[T1]{fontenc}
\usepackage{amsmath,amsfonts}
\usepackage{booktabs}
\usepackage{graphicx}
\usepackage{adjustbox}
\usepackage{float}
\usepackage{microtype}
\usepackage{xcolor}
\usepackage[most]{tcolorbox}
\usepackage{caption}
\usepackage{hyperref}
\usepackage{url}
\hypersetup{
  hidelinks,
  pdftitle={Do LLM Agents Compromise Like Humans? Risk and Responsibility in Dyadic Decision-Making},
  pdfauthor={Noah S. Okada, Ketika Garg, Wenning Deng, Dean Mobbs}
}
\makeatletter
\renewcommand{\@noticestring}{Accepted at SocialAgent: Second Workshop on Large Language Models for Social Reasoning and Simulation, NeurIPS 2026.}
\makeatother

\title{Do LLM Agents Compromise Like Humans? Risk and Responsibility in Dyadic Decision-Making}

\author{%
  Noah S.~Okada\textsuperscript{1,*} \quad
  Ketika Garg\textsuperscript{1,*} \quad
  Wenning Deng\textsuperscript{1,*} \quad
  Dean Mobbs\textsuperscript{1,2} \\[0.25em]
  {\small\textsuperscript{*}Equal contribution} \\[0.5em]
  \textsuperscript{1}Division of the Humanities and Social Sciences \\
  \textsuperscript{2}Computation and Neural Systems Program \\
  California Institute of Technology, Pasadena, CA 91125
}

\begin{document}
\raggedbottom
\maketitle

\input{sections/abstract}
\input{sections/introduction}
\input{sections/methods}
\input{sections/results}
\input{sections/discussion}

\bibliographystyle{plainnat}
\bibliography{references}

\clearpage
\appendix
\input{appendix/broader_impacts}
\input{appendix/statistical_results}

\input{appendix/computational_models}

\input{appendix/persona}

\input{appendix/prompts}
\input{appendix/data_availability}

\end{document}

%% file: sections/abstract.tex
\begin{abstract}

Large language model (LLM)-based agents are increasingly deployed in settings where they must collaborate to make joint decisions. We investigate how this collaboration unfolds under risk, asking whether agents adjust their choices to accommodate their partners, how they assign responsibility for shared outcomes, and whether responsibility attribution shapes subsequent adjustment. We designed a series of agents with systematically varying risk preferences and paired them in a repeated joint decision-making task previously conducted with humans. Like humans, agents converged toward their partners' choices. However, similar aggregate behavior arose from different dynamics: agents showed little evidence of reciprocal adjustment, tended to weigh rewards more, and their responsibility attribution was decoupled from compromise. Similarly, while they showed bias in their responsibility attribution, the outcome's valence shaped the attribution differently than in humans. Our results reveal important differences between human--human and agent--agent joint decision-making, demonstrating that similar social outcomes can emerge from different social dynamics and reasoning.
\end{abstract}

%% file: sections/introduction.tex
\section{Introduction}

Large language models (LLMs) are increasingly deployed as autonomous agents whose decisions are coupled to those of other agents or humans. Yet it remains unclear how LLM agents adapt when their preferences differ from those of a partner, particularly when joint decisions involve risk. Here, we focus on two central features of social decision-making: whether and how much agents \emph{compromise} toward a partner with a different tolerance for risk, and how they \emph{attribute responsibility} for the outcomes of their joint decisions.

Economic games have become a standard testbed for LLM social behavior. LLMs often match or exceed human cooperation, retaliate less, and favor fairness, reciprocity, and social welfare over pure self-interest \citep{fontana_nicer_2025,leng_llm_2024}. Yet similar cooperative outcomes need not reflect similar interpersonal dynamics and social reasoning. Many existing paradigms almost exclusively use binary or discrete choices over fixed payoff matrices, which cannot capture \emph{graded} mutual co-adaptation---how far an agent shifts toward a partner's preference in a continuous action space. Likewise, studies of LLM responsibility attribution rely on static, third-person vignettes \citep{bigelow2026actual}, in which models judge hypothetical actors rather than outcomes they helped produce. However, first-person attributions matter because perceived responsibility may shape whether an agent corrects course, defers, or apologizes. Yet whether LLM agents show self-serving or outcome-dependent attributions for their own actions remains unknown.

We adapt a dyadic risky-foraging paradigm, recently validated in a large human sample \citep{deng2026linking}, to LLM--LLM interaction. Across repeated trials, agents steer a joint decision in a continuous action space and share responsibility for its outcome; humans in this task compromise toward one another and show egocentric bias when attributing responsibility for shared outcomes. We ask: (1) Do LLM dyads compromise toward their partner's risk preference, and does compromise emerge through reciprocal, outcome-sensitive adjustment as in humans? (2) How do LLM agents assign responsibility for shared outcomes, and do they claim more credit for wins than blame
for losses? (3) Are these responsibility judgments and egocentric biases coupled to the tendency to compromise?

%% file: sections/methods.tex
\section{Methods}

\textbf{Dyadic risky-foraging task.} LLM agents completed the same task as the human participants in \citet{deng2026linking} (Fig.~\ref{fig:compromise}A). Agents chose among 9 locations on a risk--reward gradient: reward was known and increased quadratically with location, while the risk of predator attack, which incurred a fixed loss, also increased with location and had to be learned through experience. Two predators (low- and high-threat) imposed different risk functions. Agents first completed 120 individual trials (60 per predator), then 60 dyadic trials (30 per predator) in which both members predicted their partner's choice and then chose independently. The pair's joint location was the midpoint of the two choices, and rewards and losses were doubled and shared equally. After each outcome, agents allocated responsibility between self and partner on a 5-level 0--1 scale, framed as credit after a harvest and blame after a capture. We benchmarked agents against the human sample ($N=764$) using identical analysis code and behavioral definitions (prompts and task description in Appendix~\ref{app:Prompts}).

\textbf{Risk personas.} All agents were Llama-3.3-70B-Instruct (Appendix~\ref{app:Prompts}). To induce heterogeneous risk preference across dyads, we injected risky personas using a theory-of-planned-behavior (TPB) based persona instrument (Appendix~\ref{app:persona}). TPB holds that behavior follows from intentions shaped by attitudes, social norms, and perceived control \citep{TPB}; accordingly, each persona states the agent's attitude toward risk and intention to seek or avoid it \citep{kocielnik2026psychometric}. We validated that risk preferences rose monotonically across fitted risk curvature estimated from individual foraging choices in 8 persona variants. From these personas we retained three well-separated profiles, \emph{averse} ($\theta = 0.33$), \emph{neutral} ($0.70$), and \emph{risky} ($1.40$), for the dyadic experiment (Fig.~\ref{fig:compromise}D, Appendix~\ref{app:persona}).  Dyads paired personas within and across risk types (risky--risky, risky--neutral, risky--averse, averse--neutral; 100 dyads per composition, for a total of 800 agents).

%% file: sections/results.tex
\section{Results}
\label{sec:results}
 \subsection{Agents reproduce human-like compromise without human-like reciprocity}

\begin{figure}[H]
  \centering
  \includegraphics[width=\linewidth]{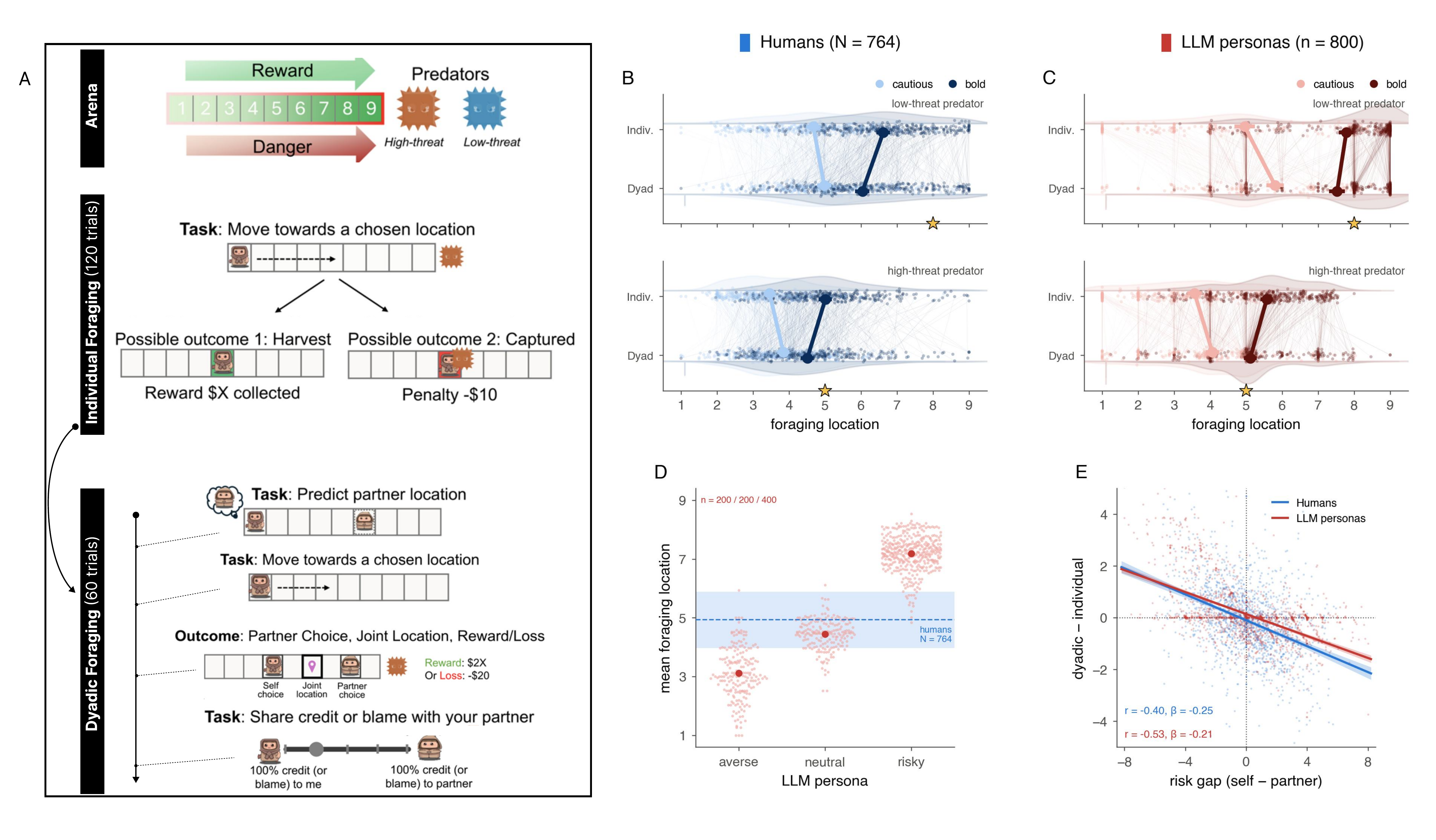}
  \caption{\textbf{Dyadic compromise and adjustment in humans and LLM agents.} \textbf{(A)} Participants foraged along a nine-step risk--reward gradient, first individually and then with a partner. On dyadic trials, participants predicted their partner's choice before choosing their own location. Outcomes depended on the partners' mean choice and were followed by responsibility attribution. \textbf{(B--C)} Mean individual and dyadic locations of cautious (light) and bold (dark) members in humans ($N=764$; B) and agents ($n=800$; C), by predator. Lines connect phases, points show role means, and stars mark risk-neutral optima. \textbf{(D)} Individual-phase mean location by persona relative to the human range. \textbf{(E)} Individual-to-dyadic change by baseline self--partner gap in humans and agents. Lines show OLS fits with 95\% CIs and Pearson r.}
  \label{fig:compromise}
\end{figure}

We first asked whether agents adjusted their risk preferences based on their partner's behavior. For each agent, we compared the average location between the individual and dyadic phases, classifying agents as the more cautious or bold member of each pair based on their individual-phase behavior. Across all dyads, both members converged toward each other, mirroring the behavior displayed by humans (Fig.~\ref{fig:compromise}B-C), with cautious agents advancing and bold agents retreating. Unlike humans, however, cautious agents shifted more than bold agents, reversing which partner made the greater compromise ($\beta=-0.503 \pm 0.103$, $z=-4.89$, $p<0.001$). Pairing increased rewards in both populations, with agents gaining more from pairing than humans (Appendix ~\ref{app:stats}).

Following \citet{deng2026linking}, we fit three dyadic choice models per agent by maximum likelihood --- \emph{as-if-individual} (partner ignored), \emph{social influence}, and \emph{other-regarding preference}, the last weighting the partner's outcome by a social weight $w$ alongside risk curvature $\theta$ and learning rate $\alpha$ (Appendix~\ref{app:computationalmodels}). As in humans, the other-regarding preference model best fit agents' dyadic behavior. However, it was preferred for fewer agents than humans (53\% vs. 68\%), and agents placed less weight on their partners ($w=0.21$ vs. $0.34$), with 42\% estimated at $w\leq0$ (Fig.~\ref{fig:blame}C). Greater partner weight predicted tighter convergence, suggesting that these estimates captured meaningful variation in social influence (Appendix ~\ref{app:stats}). These models capture overall adjustments and partner's influence, but not how that influence unfolded over time.

We also tested whether agents compromised through reciprocal adjustments by modeling each member's change in location at a trial, \textit{t}, as a function of its current distance from the partner and its partner's previous step change (Table~\ref{tab:supp_B_stepchange}). A reciprocally responsive player should adjust more when farther from its partner and respond in the complementary direction when the partner has just moved toward them. Humans showed both distance-dependent alignment ($\beta=-0.371\pm0.022$) and reciprocity ($\beta=-0.147\pm0.023$). Agents, by contrast, responded only weakly to self--partner distance ($\beta=-0.076\pm0.005$) and showed no statistically detectable reciprocity ($\beta=-0.016\pm0.010$). Thus, agents incorporated information about their partner's preferences, but less consistently and without the reciprocal adjustment observed in humans.

\subsection{Agents exhibit less egocentric bias and weaker outcome-dependent modulation in responsibility attribution}

\begin{figure}[H]
  \centering
  \includegraphics[width=\linewidth]{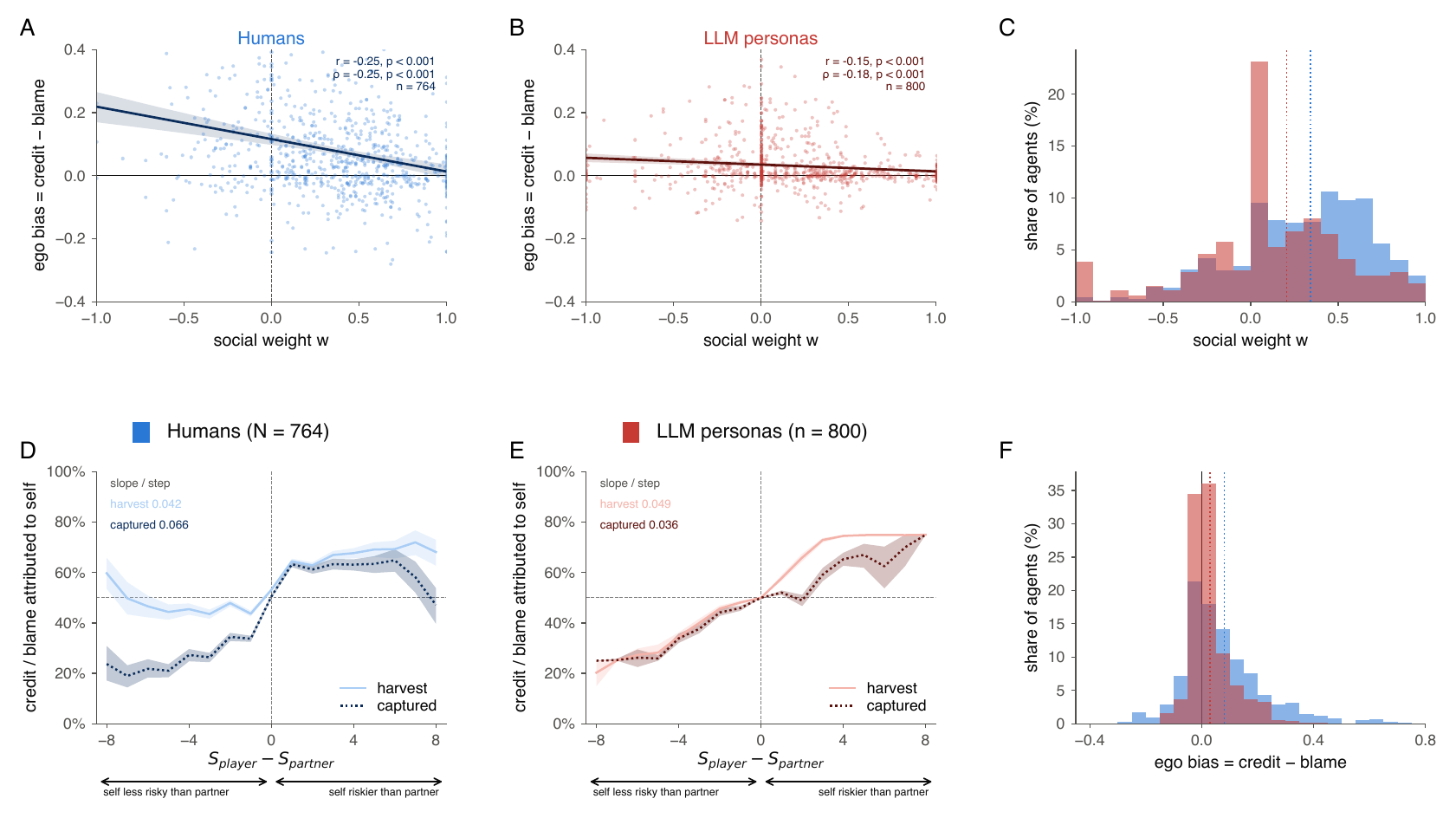}
  \caption{\textbf{Responsibility attribution and compromise.} \textbf{(A--C)} Egocentric bias versus fitted other-regarding weight $w$ in humans (A) and agents (B), with the distribution of $w$ (C). \textbf{(D--F)} Self-attributed responsibility by self--partner step gap and outcome in humans (D) and agents (E), with the distribution of egocentric bias (F). Harvests are light solid and captures dark dotted; bands show bootstrapped 95\% CIs, and vertical dotted lines mark population means.}
  \label{fig:blame}
\end{figure}

After each trial, agents divided credit (for a harvest) or blame (for a capture) between themselves and their partner. Humans show an egocentric asymmetry when making these judgments, claiming more credit for shared successes than blame for shared failures (mean difference $=0.081\pm0.006$). Agents reproduced this bias (mean difference $=0.031\pm0.003$, $d_z=0.43$), but to a lesser degree (Fig.~\ref{fig:blame}F). However, this bias in agents did not correspond with their tendency to compromise. In humans, participants with a larger social weight $w$ show a less egocentric bias ($r=-0.25$, $p<0.001$), reflecting a tendency to weight self relative to a partner in both choices and responsibility judgments (Fig.~\ref{fig:blame}A). This correlation was attenuated in agents ($r=-0.15$, $p<0.001$), and was not significant after controlling for the distance between partners ($r=-0.04$, $p=0.25$)  (Fig.~\ref{fig:blame}B).

We next asked whether partner's relative position (self--partner distance) affected agents' responsibility attributions. Agents attributed more responsibility to themselves when they made riskier choices than their partner ($\beta=0.049\pm0.001$), closely paralleling humans ($\beta=0.042\pm0.002$). However, agents diverged from humans in how trial outcome (capture or harvest) modulated this attribution pattern. In humans, the effect of relative position was larger after capture than after harvest ($\beta_{\mathrm{interaction}}=0.024 \pm 0.002$, $z=13.56$, $p<0.001$, slope (harvest) = $0.042$ vs slope (capture) = $0.066$): the riskier a player was, the more blame they absorbed, and the more cautious a player was, the more they deflected (Fig.~\ref{fig:blame}D). In agents the interaction reversed ($\beta_{\mathrm{interaction}}=-0.013 \pm 0.002$, $z=-7.30$, $p<0.001$, slope (harvest) = $0.049$ vs slope (capture) = $0.036$): relative position did not shape blame as strongly as in humans (Fig.~\ref{fig:blame}E).

%% file: sections/discussion.tex
\section{Discussion}

We compared LLMs' social behavior with humans' in a joint decision-making paradigm, and found important similarities and differences between the two \cite{deng2026linking}. Like humans, agents compromised toward each other's risk preferences and showed egocentric biases in responsibility attribution. In both cases, however, the effect was attenuated and driven by different factors. LLM agents placed less weight on their partner's preferences and were more likely to compromise if it increased their rewards. Unlike human pairs, they were less responsive to alignment gaps and to their partner's actions, neither of which gave rise to human-like reciprocal dynamics. These results reflect lower social values and overvaluation of self-interest documented for LLM agents in prior work \cite{akata_playing_2025}.

Similarly, we observed that LLM agents' responsibility attributions were driven less by choice differences and exhibited less egocentric bias than those of humans. Agents also weighted differences between their own and their partners' choices more heavily after successes than failures, the reverse of humans, who weighted them more heavily after failures. This is in line with prior work showing that LLMs emphasize others' actual contributions, and reason more like humans about successful outcomes than failures \cite{bigelow2026actual}, while extending to first-person judgments about jointly caused outcomes. Further, unlike humans, attributions did not predict compromise, suggesting that agents' social judgments about each other's contributions may be decoupled from their actions in joint tasks.

\textbf{Limitations.} We tested a single LLM, Llama-3.3-70B-Instruct, so how far these patterns generalize across architectures, scales, and post-training regimes remains open. Risk preferences were induced via personas, which may have led to more deterministic behavior than humans', plausibly reducing the social uncertainty that likely drives compromise and reciprocal adjustment in human dyads. Finally, we studied only LLM--LLM pairs, so whether these dynamics extend to human--agent teams remains unknown. Together, this work clarifies how LLM teams can differ from humans in joint decisions and highlights the value of non-binary, trial-by-trial paradigms for examining behavioral dynamics and heterogeneity.

%% file: appendix/broader_impacts.tex
\section{Broader Impacts}
LLM agents are increasingly deployed in settings where their decisions depend on those of other agents or humans, making it important to understand not only what collective outcomes they produce but how those outcomes arise. Our results show that aggregate humanlike behavior can mask markedly different underlying dynamics. Agents converged toward their partners much as humans did, yet showed little evidence of the reciprocal adjustment observed in human dyads. This distinction underscores the need for trial-by-trial behavioral audits that account for interaction history and partner behavior, rather than evaluating multi-agent systems solely by their aggregate outcomes.

Our findings also caution against relying on agents' self-reports for oversight. Agents' responsibility attributions were largely decoupled from their compromise behavior, suggesting that elicited judgments may not reliably reveal the processes governing their decisions. Oversight based on such reports alone could therefore provide a misleading picture of how a multi-agent system will behave. Finally, a simple persona was sufficient to systematically shift agents' risk taking. Although this sensitivity is useful for controlled evaluation, it also means that seemingly minor changes in persona framing could intentionally or inadvertently induce excessively risky behavior in high-stakes settings.

%% file: appendix/statistical_results.tex
\section{Statistical models and full results}
\label{app:stats}
This appendix presents the statistical models behind the results (\S \ref{sec:results}) and their full coefficient tables.

\subsection{Behavioral descriptives}
\label{sec:A_descriptives}

\input{tables/tab_supp_A_descriptives}

\input{tables/tab_supp_A_descriptives_by_persona}
\input{tables/tab_supp_A_descriptives_by_composition}

\subsection{Individual-phase learning and fitted risk preferences}
\label{sec:E_base}

To isolate the model's intrinsic task behavior before introducing persona-based manipulation of risk preferences, we first ran Llama-3.3-70B without any persona prompt. This base condition is Llama-3.3-70B-Instruct with the persona slot of the prompt left empty, so that the agent receives the task instructions and nothing else.

\begin{figure}[htbp]
  \centering
  \includegraphics[width=\linewidth]{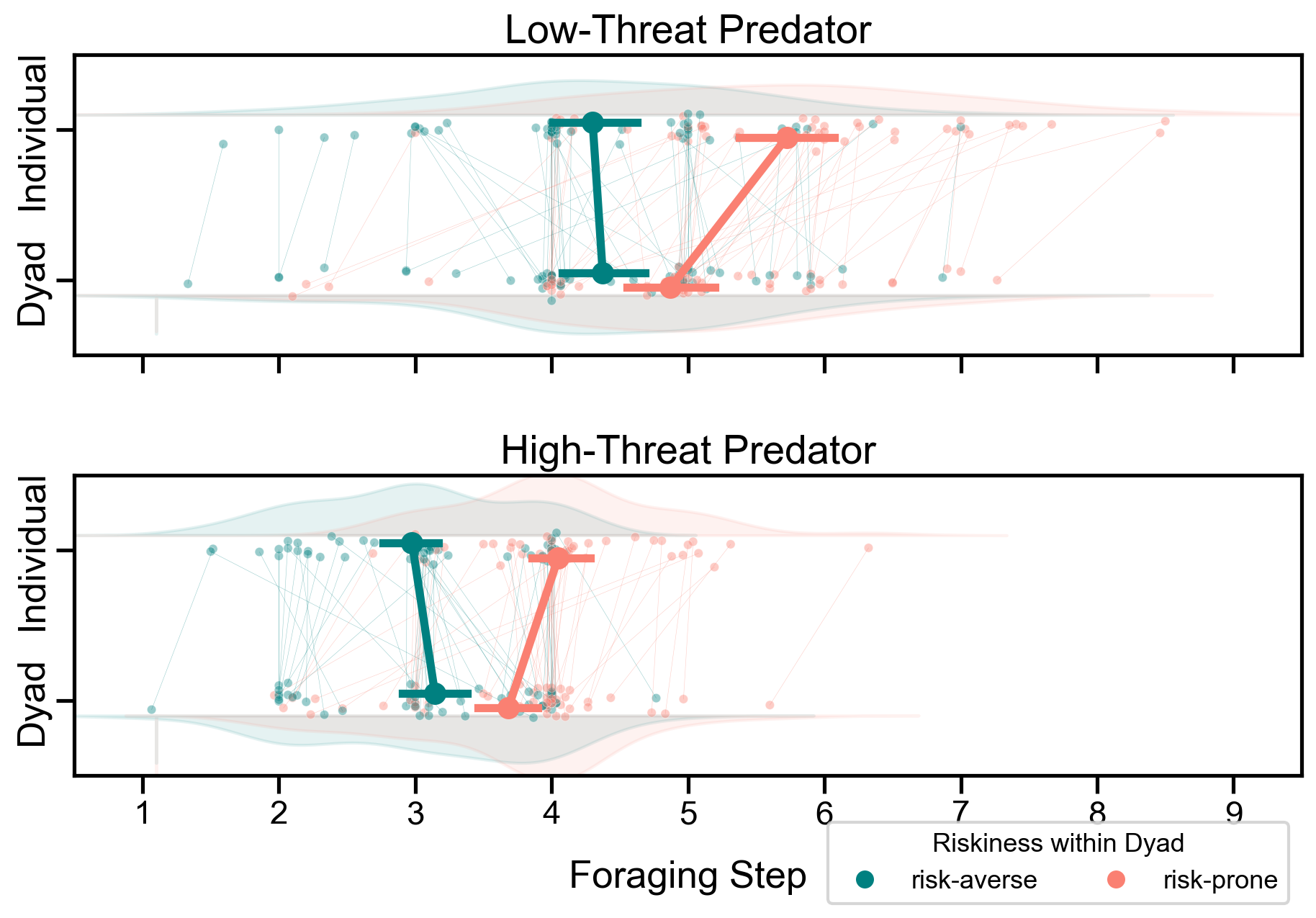}
  \caption{\textbf{Individual and dyadic foraging without persona prompts.} Mean chosen location during the final individual block (trials 61--120) and the dyadic phase for Llama-3.3-70B-Instruct agents without persona prompts (100 agents in 50 dyads), separately for low-threat and high-threat predators. Colors distinguish the more cautious (teal) and bolder (salmon) members of each dyad, classified by their individual-phase choices. Thin lines connect each agent's mean choices across phases; thick lines connect the corresponding role means.}
  \label{fig:figS_E1}
\end{figure}

Table ~\ref{tab:supp_E_base_llama}  compares agents without persona prompts with the human sample on individual-phase behavior and fitted parameters. Figure~\ref{fig:figS_E1} shows these agents' mean foraging locations across the individual and dyadic phases, separately by predator and within-dyad risk role. In the final individual block (trials 61--120), Llama-3.3-70B foraged at a mean location of 5.01 (SD 1.37) under the low-threat predator and 3.49 (SD 0.90) under the high-threat predator, against 5.66 (SD 1.70) and 4.23 (SD 1.39) for humans. The model responded to the risk gradient, but exhibited more cautious and less varied behavior compared to humans, consistently choosing safer locations across trials.

\paragraph{Llama-3.3-70B falls within the human range with less variability in behavior.} Fitting the individual-phase model to these agents gives a risk preference $\theta = 0.65$ (SD 0.19), a learning rate $\alpha = 0.71$ (SD 0.29) and a generalization parameter $\gamma = 0.46$ (SD 0.35), against $\theta = 0.74$ (SD 0.33), $\alpha = 0.53$ (SD 0.34) and $\gamma = 0.58$ (SD 0.38) in humans. Llama-3.3-70B is therefore not an outlier on any single parameter, but the spread of $\theta$ (risk preference) across its agents is roughly half the human one. This compression motivates inducing heterogeneity with personas (Appendix ~\ref{app:persona}), as the dyadic phenomena of interest are driven by the difference in risk preference between partners.

\input{tables/tab_supp_E_base_llama}

\subsection{Compromise by role}
\label{sec:AB_compromise}
Within each dyad we labeled the member with the lower mean location in the final individual block (trials 61--120) as cautious and the other as bold, and regressed each member's movement toward its partner (dyadic minus individual-phase mean location) on role, population, and their interaction, with standard errors clustered by dyad (Table~\ref{tab:supp_B_compromise}). Human bold members compromised more than cautious members, whereas Llama-3.3-70B\textsubscript{Persona} cautious members compromised more than bold members ($\beta=-0.503\pm0.103$, $z=-4.89$, $p<0.001$).
\input{tables/tab_supp_B_compromise}

\subsection{Reward gain from pairing}
\label{sec:AB_reward}
\begin{figure}[H]
  \centering
  \includegraphics[width=.45\linewidth]{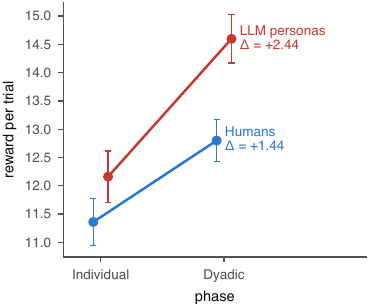}
  \caption{\textbf{Reward per trial in the individual and dyadic phases.} Mean reward per trial over agents ($\pm$95\% CI) in the individual phase (trials 61--120) and the dyadic phase, for humans ($N=764$) and Llama-3.3-70B\textsubscript{Persona} ($n=800$); $\Delta$ is the within-agent gain from pairing.}
  \label{fig:supp_AB_reward}
\end{figure}

For each member and predator with a complete dyad we computed the reward gain from pairing, half the joint dyadic reward minus the member's own individual-phase reward, and regressed it on population, with standard errors clustered by dyad within predator (Figure~\ref{fig:supp_AB_reward}). Pairing increased reward in both populations (humans $+1.44$ points per trial, $t=8.14$; agents $+2.44$, $t=12.57$; both $p<0.001$), with agents gaining more from pairing than humans ($\beta=+1.00\pm0.29$, $z=3.39$, $p=0.001$).

\input{tables/tab_reward_gain}

\subsection{Step change and responsibility attribution}
\label{sec:AB_models}
\paragraph{Step-change model.} Following Deng et al.~\citep{deng2026linking}, we modeled each member's change in location on the next trial as a function of its current distance from its partner and of its partner's previous step change,
\begin{equation}
\Delta x_{i,t+1} = \beta_0 + \beta_1\, d_{i,t} + \beta_2\, c_t + \beta_3\, d_{i,t} c_t + \beta_4\, \Delta x_{j,t} + u_i + \varepsilon_{i,t},
\label{eq:B_stepchange}
\end{equation}
where $d_{i,t}$ is member $i$'s own minus its partner's location on trial $t$, $c_t$ indicates a capture (reference: harvest), $\Delta x_{j,t}$ is the partner's previous step change, and $u_i$ collects a random intercept and random slopes per agent. The model was fit within each population and then pooled with a population indicator (Table~\ref{tab:supp_B_stepchange}). Agents responded less to distance than humans and not at all to their partner's previous step (distance-by-population interaction $\beta=0.331\pm0.013$; previous-step-by-population interaction $\beta=0.148\pm0.014$; both $p<0.001$).

\paragraph{Responsibility-attribution model.} We modeled the share of the outcome a member attributed to itself as a function of its relative position, the trial outcome, and their interaction,
\begin{equation}
s_{i,t} = \beta_0 + \beta_1\, g_{i,t} + \beta_2\, c_t + \beta_3\, g_{i,t} c_t + u_i + \varepsilon_{i,t},
\label{eq:B_attribution}
\end{equation}
where $s_{i,t}$ is the mean-centered self-attributed share (0--1), $g_{i,t}$ is the member's own minus its partner's location, and $c_t$ and $u_i$ are as in Eq.~\ref{eq:B_stepchange}; because capture is coded against a harvest reference, $\beta_3$ is the change in the contribution slope after a capture. The model was fit within each population and then pooled (Table~\ref{tab:supp_B_attribution}). Capture strengthened the contribution slope in humans but weakened it in agents (three-way interaction of position, capture and population, $\beta=-0.039\pm0.003$, $p<0.001$).
\input{tables/tab_supp_B_stepchange}
\input{tables/tab_supp_B_attribution}

\subsection{Egocentric bias and use of the attribution scale}
\label{sec:AB_egobias}
\begin{figure}[htbp]
  \centering
  \includegraphics[width=\linewidth]{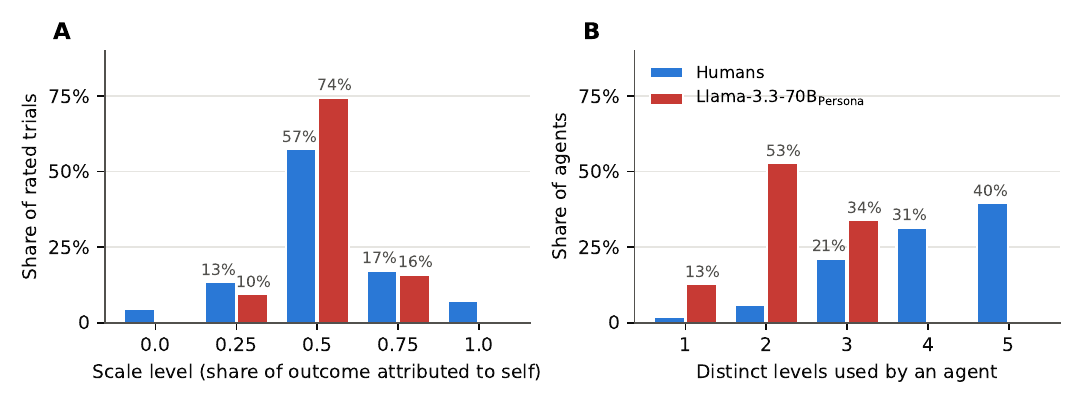}
  \caption{\textbf{Use of the five-level responsibility scale.} (A) Share of rated dyadic trials at each level of the self-attributed share (0 = entirely the partner, 1 = entirely oneself), for humans ($n = 764$; 44,899 trials) and Llama-3.3-70B\textsubscript{Persona} ($n = 800$; 48,000 trials). (B) Share of agents by the number of distinct levels they used across their 60 dyadic trials. Agents answered at the midpoint on 74\% of trials and never used the top level, whereas humans spread their ratings across all five levels.}
  \label{fig:supp_B_scale_usage}
\end{figure}

For each agent we computed mean self-credit after a harvest minus mean self-blame after a capture, tested it against zero within each population, compared the populations with a cluster-robust contrast, and split the per-agent means by the agent's position relative to its partner on the trial; Figure~\ref{fig:supp_B_scale_usage} shows how each population used the five-level scale (Tables~\ref{tab:supp_B_egobias},\ref{tab:supp_AB_ego_position}). Agents showed a smaller bias than humans ($\beta=-0.050\pm0.007$, $z=-7.63$, $p<0.001$) and used a narrower range of the scale ($2.2$ vs.\ $4.0$ distinct levels per agent; 74\,\% of trials at the midpoint, none at the top level), and what bias they showed sat in the opposite member: in humans it is carried by the safer partner, who accepts far less than half the blame after a capture (self-credit $0.469$ vs.\ self-blame $0.300$), whereas the safer agent split credit and blame nearly identically ($0.434$ vs.\ $0.430$) and only the riskier agent accepted somewhat less blame than the credit it claimed ($0.656$ vs.\ $0.568$).
\input{tables/tab_supp_B_egobias}
\input{tables/tab_supp_AB_ego_position}
\subsection{Egocentric bias and social weight}
\label{sec:AB_egow}
Following Deng et al.~\citep{deng2026linking}, we correlated each agent's egocentric bias with its fitted social weight $w$, partialled out the mean distance between partners, and regressed the bias on both members' weights (Table~\ref{tab:egobias_w}); to compare the populations we pooled them on within-population standardized variables with population-by-weight interactions (Table~\ref{tab:supp_AB_egobias_w}). The association was negative in both populations but weaker in agents ($r=-0.25$ vs.\ $-0.15$), did not survive the distance control in agents (partial $r=-0.04$, $p=0.25$; humans $-0.13$, $p<0.001$), and on standardized scales the dependence on an agent's own weight was less than half the human one ($\beta=-0.09\pm0.04$ vs.\ $-0.23\pm0.04$; interaction $p=0.011$), whereas the dependence on the partner's weight did not differ ($p=0.38$).
\input{tables/tab_egobias_w}
\input{tables/tab_supp_AB_egobias_w}

%% file: tables/tab_supp_A_descriptives.tex
\begin{table}[htbp]
\centering
\small
\caption{\textbf{Task behavior across populations.} Mean chosen location (1--9), mean reward per trial and capture rate on solo trials and on the dyadic trials, for humans, Llama-3.3-70B and Llama-3.3-70B\textsubscript{Persona}. Values are means across agents; dyadic reward is each agent's share of the doubled joint payoff and the joint location is the midpoint the pair foraged at. Llama-3.3-70B\textsubscript{Persona} forages further forward than humans and is captured more often, whereas Llama-3.3-70B is the most conservative of the three. }
\label{tab:supp_A_descriptives}
\begin{tabular*}{\textwidth}{@{\extracolsep{\fill}}l cc cc cc@{}}
\toprule
& \multicolumn{2}{c}{Humans}
& \multicolumn{2}{c}{Llama-3.3-70B}
& \multicolumn{2}{c}{Llama-3.3-70B\textsubscript{Persona}} \\
\cmidrule(lr){2-3} \cmidrule(lr){4-5} \cmidrule(l){6-7}
& \multicolumn{2}{c}{$n = 764$}
& \multicolumn{2}{c}{$n = 100$}
& \multicolumn{2}{c}{$n = 800$} \\
Measure & $M$ & $SD$ & $M$ & $SD$ & $M$ & $SD$ \\
\midrule
\multicolumn{7}{@{}l}{\itshape Solo (trials 61--120)} \\
\quad Chosen location (1--9)   & 4.94  & 1.36  & 4.26  & 0.70  & 5.48  & 1.89 \\
\quad Reward per trial          & 11.40 & 4.12  & 11.62 & 3.62  & 12.29 & 4.80 \\
\quad Capture rate              & 0.327 & 0.160 & 0.238 & 0.078 & 0.394 & 0.212 \\
\addlinespace
\multicolumn{7}{@{}l}{\itshape Dyadic (60 trials)} \\
\quad Chosen location (1--9)    & 4.85  & 1.23  & 4.01  & 0.64  & 5.63  & 1.44 \\
\quad Joint location (midpoint) & 5.06  & 0.93  & 4.19  & 0.61  & 5.80  & 1.13 \\
\quad Reward per trial (own share) & 12.77 & 3.76 & 11.40 & 3.22 & 14.71 & 3.85 \\
\quad Capture rate              & 0.341 & 0.113 & 0.233 & 0.070 & 0.417 & 0.137 \\
\bottomrule
\end{tabular*}
\end{table}

%% file: tables/tab_supp_A_descriptives_by_persona.tex
\begin{table}[htbp]
\centering
\small
\caption{\textbf{Behavior during the individual phase by prompt risk profile.} The individual-trial measures of Table~\ref{tab:supp_A_descriptives} for Llama-3.3-70B\textsubscript{Persona}, split by the risk profile in the prompt (averse, neutral, risky; $n$ counts agents).}
\label{tab:supp_A_descriptives_by_persona}
\begin{tabular*}{\textwidth}{@{\extracolsep{\fill}}lccc@{}}
\toprule
Measure & Averse & Neutral & Risky \\
$n$ agents & 200 & 200 & 400 \\
\midrule
\multicolumn{4}{@{}l}{\emph{Solo (trials 61--120)}} \\
\quad Mean chosen location (1--9) & 3.11 (0.88) & 4.45 (0.52) & 7.18 (0.60) \\
\quad Mean reward per trial & 7.70 (3.38) & 12.09 (2.78) & 14.68 (4.51) \\
\quad Capture rate & 0.145 (0.089) & 0.266 (0.067) & 0.581 (0.098) \\
\bottomrule
\end{tabular*}
\end{table}

%% file: tables/tab_supp_A_descriptives_by_composition.tex
\begin{table}[htbp]
\centering
\small
\caption{\textbf{Behavior during the dyadic phase by dyad composition.} The dyadic-trial measures of Table~\ref{tab:supp_A_descriptives} for Llama-3.3-70B\textsubscript{Persona}, split by the four persona pairings (100 dyads, 200 agents each). Mean chosen location is averaged over both members. Reward per trial rises with the pair's risk appetite, and the mixed pairings forage at a joint location between their members' solo positions.}
\label{tab:supp_A_descriptives_by_composition}
\begin{tabular}{lcccc}
\toprule
Measure & Risky--risky & Risky--neutral & Risky--averse & Averse--neutral \\
$n$ agents & 200 & 200 & 200 & 200 \\
\midrule
\multicolumn{5}{l}{\emph{Dyadic (60 trials)}} \\
\quad Mean chosen location (1--9) & 6.93 (0.63) & 5.76 (1.26) & 5.67 (1.38) & 4.14 (0.68) \\
\quad Mean joint location (midpoint) & 7.07 (0.51) & 5.98 (0.55) & 5.86 (0.57) & 4.30 (0.56) \\
\quad Mean reward per trial (own share) & 15.69 (4.09) & 15.79 (3.25) & 15.86 (3.35) & 11.48 (2.66) \\
\quad Capture rate & 0.560 (0.088) & 0.437 (0.081) & 0.417 (0.085) & 0.255 (0.082) \\
\bottomrule
\end{tabular}
\end{table}

%% file: tables/tab_supp_E_base_llama.tex
\begin{table}[htbp]
\centering
\footnotesize
\caption{\textbf{Llama-3.3-70B, against the human sample.} Individual-phase foraging location by predator and the absolute within-dyad difference in baseline location, mean (SD) over agents; the three parameters of the individual-phase model, mean (SD) of the per-agent maximum-likelihood fit; and the per-agent egocentric bias regressed on the agent's own fitted social weight $w$ and its partner's, estimate with stars and 95\% confidence interval (Llama-3.3-70B $n = 100$ agents in 50 dyads; humans $n = 764$ agents). $^{*}p < .05$, $^{**}p < .01$, $^{***}p < .001$; ns = not significant).}
\label{tab:supp_E_base_llama}
\begin{tabular*}{\textwidth}{@{\extracolsep{\fill}}lcc@{}}
\toprule
\textbf{Quantity} & \textbf{Llama-3.3-70B} & \textbf{Humans} \\
\midrule
\multicolumn{3}{@{}l}{Individual phase: location, mean (SD) over agents} \\
Low-threat predator & 5.01 (1.37) & 5.66 (1.70) \\
High-threat predator & 3.49 (0.90) & 4.23 (1.39) \\
Within-dyad $|$baseline gap$|$ & 1.21 (1.02), median 0.96 & 1.71 (1.31), median 1.40 \\
\addlinespace[2pt]
\multicolumn{3}{@{}l}{Individual-phase model, mean (SD) of the per-agent fit} \\
$\theta$ (risk preference) & 0.65 (0.19) & 0.74 (0.33) \\
$\alpha$ (learning rate) & 0.71 (0.29) & 0.53 (0.34) \\
$\gamma$ (generalization) & 0.46 (0.35) & 0.58 (0.38) \\
\addlinespace[2pt]
\multicolumn{3}{@{}l}{Egocentric bias regressed on own and partner's social weight, $\beta$ [95\% CI]} \\
Own $w$ & $+0.001$ ns {\scriptsize $[-0.041, +0.044]$} & $-0.095$*** {\scriptsize $[-0.124, -0.066]$} \\
Partner's $w$ & $-0.078$*** {\scriptsize $[-0.120, -0.035]$} & $-0.031$* {\scriptsize $[-0.060, -0.001]$} \\
\bottomrule
\end{tabular*}
\end{table}

%% file: tables/tab_supp_B_compromise.tex
\begin{table}[htbp]
\centering
\small
\caption{\textbf{Compromise by role and population.} Movement toward the partner (dyadic minus individual-phase mean location, signed toward the partner) regressed on member role, population and their interaction; OLS with standard errors clustered by dyad.}
\label{tab:supp_B_compromise}
\begin{tabular}{lrrrr}
\toprule
Term & $\beta$ & SE & $z$ & $p$ \\
\midrule
Intercept (humans, cautious) & $+0.344$ & 0.055 & 6.28 & $<$0.001 \\
Bold member & $+0.196$ & 0.086 & 2.29 & 0.022 \\
Llama-3.3-70B\textsubscript{Persona} & $+0.322$ & 0.074 & 4.33 & $<$0.001 \\
Bold member $\times$ Llama-3.3-70B\textsubscript{Persona} & $-0.503$ & 0.103 & -4.89 & $<$0.001 \\
\bottomrule
\end{tabular}
\end{table}

%% file: tables/tab_reward_gain.tex
\begin{table}[htbp]
\centering
\footnotesize
\caption{\textbf{Reward gain from pairing by population}. Gain was defined as half the joint dyadic reward minus the member's individual-phase reward. Population differences were estimated by OLS (human reference), with standard errors clustered by dyad--predator cell ($n=3{,}056$ member--predator cells).
}
\label{tab:reward_gain}
\begin{tabular}{lrrrrc}
\toprule
\textbf{Term} & $\beta$ & \textbf{SE} & $z$ & $p$ & \textbf{95\% CI} \\
\midrule
Intercept (humans) & $+1.442$ & 0.199 & 7.25 & $<$0.001 & [1.052, 1.831] \\
Llama-3.3-70B\textsubscript{Persona} & $+0.997$ & 0.294 & 3.39 & 0.001 & [0.421, 1.573] \\
\bottomrule
\end{tabular}
\end{table}

%% file: tables/tab_supp_B_stepchange.tex
\begin{table}[htbp]
\centering
\small
\caption{\textbf{Step-change model (Eq.~\ref{eq:B_stepchange}).} Fixed effects, fit separately to each population and then pooled with a population indicator; every $\times$ Llama-3.3-70B\textsubscript{Persona} term is the difference from humans.}
\label{tab:supp_B_stepchange}
\begin{tabular}{lrrrr}
\toprule
Term & $\beta$ & SE & $z$ & $p$ \\
\midrule
\multicolumn{5}{l}{Humans ($n = 764$ participants, 41,454 trials)} \\
\midrule
Intercept & $+0.172$ & 0.012 & 14.86 & $<$0.001 \\
Capture & $-0.357$ & 0.018 & -19.33 & $<$0.001 \\
Distance from partner & $-0.371$ & 0.022 & -16.96 & $<$0.001 \\
Distance $\times$ capture & $-0.026$ & 0.021 & -1.21 & 0.228 \\
Partner's previous step & $-0.147$ & 0.023 & -6.47 & $<$0.001 \\
\midrule
\multicolumn{5}{l}{Llama-3.3-70B\textsubscript{Persona} ($n = 800$ agents, 44,800 trials)} \\
\midrule
Intercept & $-0.001$ & 0.006 & -0.10 & 0.924 \\
Capture & $-0.038$ & 0.006 & -6.36 & $<$0.001 \\
Distance from partner & $-0.076$ & 0.005 & -14.88 & $<$0.001 \\
Distance $\times$ capture & $-0.043$ & 0.006 & -7.39 & $<$0.001 \\
Partner's previous step & $-0.016$ & 0.010 & -1.63 & 0.104 \\
\midrule
\multicolumn{5}{l}{Pooled ($n = 1{,}564$ agents, 86,254 trials)} \\
\midrule
Intercept & $+0.186$ & 0.015 & 12.71 & $<$0.001 \\
Capture & $-0.388$ & 0.019 & -19.99 & $<$0.001 \\
Llama-3.3-70B\textsubscript{Persona} & $-0.179$ & 0.021 & -8.50 & $<$0.001 \\
Capture $\times$ Llama-3.3-70B\textsubscript{Persona} & $+0.346$ & 0.027 & 12.85 & $<$0.001 \\
Distance from partner & $-0.403$ & 0.009 & -45.07 & $<$0.001 \\
Distance $\times$ capture & $-0.002$ & 0.008 & -0.30 & 0.763 \\
Distance $\times$ Llama-3.3-70B\textsubscript{Persona} & $+0.331$ & 0.013 & 25.12 & $<$0.001 \\
Distance $\times$ capture $\times$ Llama-3.3-70B\textsubscript{Persona} & $-0.020$ & 0.011 & -1.74 & 0.083 \\
Partner's previous step & $-0.172$ & 0.006 & -26.84 & $<$0.001 \\
Partner's previous step $\times$ Llama-3.3-70B\textsubscript{Persona} & $+0.148$ & 0.014 & 10.37 & $<$0.001 \\
\bottomrule
\end{tabular}
\end{table}

%% file: tables/tab_supp_B_attribution.tex
\begin{table}[htbp]
\centering
\small
\caption{\textbf{Responsibility-attribution model (Eq.~\ref{eq:B_attribution}).} Fixed effects, fit separately to each population and then pooled with a population indicator; every $\times$ Llama-3.3-70B\textsubscript{Persona} term is the difference from humans.}
\label{tab:supp_B_attribution}
\begin{tabular}{lrrrr}
\toprule
Term & $\beta$ & SE & $z$ & $p$ \\
\midrule
\multicolumn{5}{l}{Humans ($n = 764$ participants, 44,899 trials)} \\
\midrule
Intercept & $+0.041$ & 0.003 & 12.70 & $<$0.001 \\
Capture & $-0.061$ & 0.005 & -11.25 & $<$0.001 \\
Relative position & $+0.042$ & 0.002 & 24.93 & $<$0.001 \\
Relative position $\times$ capture & $+0.024$ & 0.002 & 13.56 & $<$0.001 \\
\midrule
\multicolumn{5}{l}{Llama-3.3-70B\textsubscript{Persona} ($n = 800$ agents, 48,000 trials)} \\
\midrule
Intercept & $+0.014$ & 0.002 & 6.17 & $<$0.001 \\
Capture & $-0.015$ & 0.002 & -7.73 & $<$0.001 \\
Relative position & $+0.049$ & 0.001 & 33.87 & $<$0.001 \\
Relative position $\times$ capture & $-0.013$ & 0.002 & -7.30 & $<$0.001 \\
\midrule
\multicolumn{5}{l}{Pooled ($n = 1{,}564$ agents, 92,899 trials)} \\
\midrule
Intercept & $+0.041$ & 0.003 & 15.77 & $<$0.001 \\
Capture & $-0.059$ & 0.004 & -14.48 & $<$0.001 \\
Llama-3.3-70B\textsubscript{Persona} & $-0.026$ & 0.004 & -6.89 & $<$0.001 \\
Capture $\times$ Llama-3.3-70B\textsubscript{Persona} & $+0.039$ & 0.006 & 6.65 & $<$0.001 \\
Relative position & $+0.042$ & 0.001 & 28.63 & $<$0.001 \\
Relative position $\times$ capture & $+0.026$ & 0.002 & 15.12 & $<$0.001 \\
Relative position $\times$ Llama-3.3-70B\textsubscript{Persona} & $+0.005$ & 0.002 & 2.35 & 0.019 \\
Relative position $\times$ capture $\times$ Llama-3.3-70B\textsubscript{Persona} & $-0.039$ & 0.003 & -15.12 & $<$0.001 \\
\bottomrule
\end{tabular}
\end{table}

%% file: tables/tab_supp_B_egobias.tex
\begin{table}[htbp]
\centering
\small
\caption{\textbf{Egocentric bias by population.} Mean self-credit after a harvest minus mean self-blame after a capture, per agent; paired $t$-tests within each population and an OLS contrast between them with standard errors clustered by dyad.}
\label{tab:supp_B_egobias}
\begin{tabular}{lrrrr}
\toprule
Term & Mean & SE & Test & $p$ \\
\midrule
Humans ($n = 764$) & $+0.081$ & 0.006 & $t = 13.26$, $d_z = 0.48$ & $<$0.001 \\
Llama-3.3-70B\textsubscript{Persona} ($n = 800$) & $+0.031$ & 0.003 & $t = 12.10$, $d_z = 0.43$ & $<$0.001 \\
Llama-3.3-70B\textsubscript{Persona} $-$ humans & $-0.050$ & 0.007 & $z = -7.63$ & $<$0.001 \\
\bottomrule
\end{tabular}
\end{table}

%% file: tables/tab_supp_AB_ego_position.tex
\begin{table}[htbp]
\centering
\footnotesize
\caption{\textbf{Credit and blame by relative position.} Mean share of the outcome attributed to self after a harvest (credit) and after a capture (blame), by whether the member sat at a safer, equal or riskier location than its partner on that trial; the difference is the per-agent paired mean of credit minus blame within the position class ($\pm$SE). Humans $n = 764$, Llama-3.3-70B\textsubscript{Persona} $n = 800$.}
\label{tab:supp_AB_ego_position}
\begin{tabular}{lccc@{\hspace{1.5em}}ccc}
\toprule
 & \multicolumn{3}{c}{\textbf{Humans}} & \multicolumn{3}{c}{\textbf{Llama-3.3-70B\textsubscript{Persona}}} \\
\cmidrule(lr){2-4}\cmidrule(lr){5-7}
\textbf{Position} & credit & blame & credit $-$ blame & credit & blame & credit $-$ blame \\
\midrule
safer than partner   & 0.469 & 0.300 & $+0.147 \pm 0.008^{***}$ & 0.434 & 0.430 & $+0.014 \pm 0.004^{***}$ \\
equal                & 0.524 & 0.501 & $+0.026 \pm 0.005^{***}$ & 0.500 & 0.500 & $0.000$ \\
riskier than partner & 0.656 & 0.619 & $+0.019 \pm 0.008^{*}$   & 0.656 & 0.568 & $+0.098 \pm 0.007^{***}$ \\
\bottomrule
\end{tabular}
\end{table}

%% file: tables/tab_egobias_w.tex
\begin{table}[htbp]
\centering
\footnotesize
\caption{\textbf{Egocentric bias by own and partner's social weight}. OLS models were fit for each population using one observation per agent and the agent's own and partner's fitted other-regarding weights $w$ as predictors. Entries report coefficients, standard errors, $t$ statistics, $p$-values and 95\% confidence intervals.}
\label{tab:egobias_w}
\begin{tabular}{lrrrrc}
\toprule
\textbf{Term} & $\beta$ & \textbf{SE} & $t$ & $p$ & \textbf{95\% CI} \\
\midrule
\addlinespace[2pt]
\multicolumn{6}{l}{\textit{Humans ($n = 764$ agents)}} \\
\addlinespace[1pt]
Intercept & $+0.124$ & 0.009 & 14.38 & $<$0.001 & [0.107, 0.141] \\
Own $w$ & $-0.095$ & 0.015 & $-6.32$ & $<$0.001 & [$-0.124$, $-0.066$] \\
Partner's $w$ & $-0.031$ & 0.015 & $-2.05$ & 0.041 & [$-0.060$, $-0.001$] \\
\addlinespace[2pt]
\multicolumn{6}{l}{\textit{Llama-3.3-70B\textsubscript{Persona} ($n = 800$ agents)}} \\
\addlinespace[1pt]
Intercept & $+0.037$ & 0.003 & 13.34 & $<$0.001 & [0.032, 0.043] \\
Own $w$ & $-0.014$ & 0.006 & $-2.39$ & 0.017 & [$-0.025$, $-0.002$] \\
Partner's $w$ & $-0.018$ & 0.006 & $-3.10$ & 0.002 & [$-0.029$, $-0.006$] \\
\bottomrule
\end{tabular}
\end{table}

%% file: tables/tab_supp_AB_egobias_w.tex
\begin{table}[htbp]
\centering
\footnotesize
\caption{\textbf{Egocentric bias and social weight.} Per-agent egocentric bias (mean self-credit after a harvest minus mean self-blame after a capture) against the agent's own fitted social weight $w$ and its partner's. Rows 1--3: Pearson correlations, raw and after partialling out the mean distance between partners; the last column tests the human--agent difference (Fisher $z$). Rows 4--5: OLS of bias on both weights, coefficients on within-population standardized variables; the last column is the population $\times$ weight interaction from the pooled model (HC3 standard errors). Humans $n = 764$, Llama-3.3-70B\textsubscript{Persona} $n = 800$.}
\label{tab:supp_AB_egobias_w}
\begin{adjustbox}{max width=\linewidth}
\begin{tabular}{lccc}
\toprule
\textbf{Quantity} & \textbf{Humans} & \textbf{Llama-3.3-70B\textsubscript{Persona}} & \textbf{Difference} \\
\midrule
$r$(bias, own $w$)                      & $-0.25^{***}$ & $-0.15^{***}$ & --- \\
partial $r$(bias, own $w \mid$ distance) & $-0.13^{***}$ & $-0.04$ ns ($p=.25$) & $p=.063$ \\
partial $r$(bias, partner's $w \mid$ distance) & $+0.00$ ns & $-0.06$ ns ($p=.10$) & $p=.22$ \\
\midrule
standardized $\beta$, own $w$            & $-0.23 \pm 0.04^{***}$ & $-0.09 \pm 0.04^{**}$ & $+0.14 \pm 0.05$, $p=.011$ \\
standardized $\beta$, partner's $w$      & $-0.07 \pm 0.04$ ($p=.06$) & $-0.12 \pm 0.04^{***}$ & $-0.05 \pm 0.05$, $p=.38$ \\
\bottomrule
\end{tabular}
\end{adjustbox}
\end{table}

%% file: appendix/computational_models.tex
\section{Computational models}
\label{app:computationalmodels}
In order to capture how an agent learns and generalizes the attack pattern of each predator, and how it uses this information to coordinate with its partner, we fit a series of computational models to the behavior. On every trial the model first updates the learned attack probability of the chosen location from the outcome, then propagates that update to the other locations, and finally chooses a location by balancing risk against reward. Each model is fit to an agent's complete behavioral data, 120 individual-phase and 60 dyadic-phase trials, using maximum likelihood estimation.

For the dyadic phase we compare three classes of models that differ only in how the partner enters the choice. The \emph{as-if-individual} model assumes that an agent ignores its partner and chooses to maximize its own utility. The \emph{social influence} model assumes that the agent's internal risk preference is shifted by its partner's risk preference. The \emph{other-regarding preference} model assumes that the agent values both its own risk preference and its partner's, combining the two with a social weight $w$.

Locations are $x\in\{1,\dots,9\}$ with reward $r(x)=x^{2}$ and a fixed loss of $10$ on an attack. For each predator an agent carries learned safety values $V(x)\in[0,1]$, the estimated probability of \emph{not} being attacked at $x$, initialized to the descending ramp $V(x)=1-(x-1)/9$. The learning rate $\alpha\in[0,1]$ scales the update of $V$; the propagation strength $\gamma\in[0,1]$ spreads that update to the other locations, decaying with distance from the chosen location; the reward-sensitivity parameter $\theta\in[0,1.5]$ controls the curvature of the subjective utility function over rewards and losses, so that $\theta<1$ is risk-averse, $\theta=1$ risk-neutral and $\theta>1$ risk-seeking; the inverse temperature $\beta=5$ is fixed rather than fitted; and the social weight $w\in[-1,1]$ and the shift in riskiness $\Delta$ enter only the dyadic models defined below. On each dyadic trial $\hat{x}_p$ denotes the agent's stated prediction of its partner's location.

\subsection{Individual learner}

\textbf{Safety learning.} After a trial at location $c$ with outcome $a\in\{0,1\}$ ($1$ = attacked), the prediction error is $\delta=(1-a)-V(c)$ and the learning rate decays with the number of encounters $t$ of that predator, $\alpha_t=\alpha/\sqrt{t}$; in the dyadic phase the encounter count continues from the individual phase, $t=60+t_{\mathrm{dyadic}}$. The update is applied at $c$ and generalized to every other location with propagation strength $\gamma$ decaying with distance in both directions,
\begin{equation}
\begin{array}{ll}
V(c) \leftarrow V(c) + \alpha_t\,\delta, & \\[2pt]
V(x) \leftarrow V(x) + \alpha_t\,\gamma^{\,|x-c|}\,\delta, & x\neq c,
\end{array}
\label{eq:D_update}
\end{equation}
with $V$ clipped to $[0,1]$ after every update.

\textbf{Location value and choice.} The value of a location is its risk-curved expected payoff,
\begin{equation}
Q(x) = V(x)\,r(x)^{\theta} - \bigl(1-V(x)\bigr)\,10^{\theta},
\label{eq:D_Q}
\end{equation}
rescaled to a self-preference vector $\tilde{Q}(x)=(Q(x)-\min_{x'}Q(x'))/(\max_{x'}Q(x')-\min_{x'}Q(x'))$ over the nine locations. The choice probability is a softmax over whichever location value $L(x)$ the model in question supplies,
\begin{equation}
P(x)=\frac{\exp\bigl(\beta\,L(x)\bigr)}{\sum_{x'}\exp\bigl(\beta\,L(x')\bigr)},
\label{eq:D_softmax}
\end{equation}
and in the individual phase $L(x)=\tilde{Q}(x)$ for every model. The safety values each predator has reached at the end of the individual phase is the starting $V$ for that predator's dyadic block, and the dyadic likelihood continues to update $V$ from the dyad's realized joint location and outcome. The two phases are fit jointly, so the objective is the summed negative log-likelihood of the individual-phase and dyadic-phase choices.

\subsection{Dyadic models}

The three models differ only in how the partner enters the location value $L(x)$ passed to Eq.~\ref{eq:D_softmax}. The partner's preference is represented by a Gaussian kernel around the agent's stated prediction,
\begin{equation}
O(x)=\exp\!\left(-\frac{(x-\hat{x}_p)^{2}}{2\sigma^{2}}\right), \qquad \sigma=1,
\label{eq:D_O}
\end{equation}
where, when no prediction was recorded, $\hat{x}_p$ is the partner's last observed location and, on the first trial, location 5.

\textbf{As-if-individual model (AI).} The as-if-individual model applies the individual learner unchanged to the dyadic phase, with three parameters $\alpha,\theta,\gamma$. An agent described by this model neither moves toward nor away from its partner beyond what the shared outcomes teach it about the predators.
\begin{equation}
L^{\mathrm{AI}}(x)=\tilde{Q}(x)
\label{eq:D_AI}
\end{equation}

\textbf{Social-influence model (SI).} The social-influence model lets the partner shift the agent's own risk attitude: in the dyadic phase the reward sensitivity becomes $\theta+\Delta$, and the location value is the self-preference recomputed with the shifted curvature, giving four parameters $\alpha,\theta,\gamma,\Delta$. A positive $\Delta$ makes the agent bolder when paired, a negative one more cautious, regardless of where the partner is predicted to go.
\begin{equation}
L^{\mathrm{SI}}(x)=\tilde{Q}(x)\big|_{\theta\,\to\,\max(\theta+\Delta,\,0)}
\label{eq:D_SI}
\end{equation}

\textbf{Other-regarding model (OR).} The other-regarding model lets the social weight $w$ arbitrate between the agent's own preference and the partner's predicted preference, giving four parameters $\alpha,\theta,\gamma,w$. Agents with larger $w$ are predicted to move further toward the location they expect their partner to choose.
\begin{equation}
L^{\mathrm{OR}}(x)=\left\{\begin{array}{ll}
w\,O(x) + (1-w)\,\tilde{Q}(x), & w\ge 0,\\[3pt]
-w\,C(x) + (1+w)\,\tilde{Q}(x), & w<0,
\end{array}\right.
\qquad
C(x)=\sum_{j:\;\mathrm{clip}(2x^{*}-j)=x} O(j),
\label{eq:D_OR}
\end{equation}
where $x^{*}=\arg\max_{x}\tilde{Q}(x)$ is the agent's own preferred location and $C$ is the partner's preference mirrored about $x^{*}$, clipped to $1\dots 9$. In other words, $w=0$ is insensitivity to the partner; $w>0$ compromises toward the partner's predicted location, with $w=1$ full alignment; and $w<0$ counteracts it, moving away from the partner by as much as the partner is expected to move toward the agent, with $w=-1$ complete counter alignment.

\subsection{Estimation \& model comparison}

Parameters were estimated per agent by maximum likelihood with a three-stage grid search. The first stage evaluated a grid of step $0.2$ over the full box $\alpha,\gamma\in[0,1]$, $\theta\in[0,1.5]$, $w\in[-1,1]$ (the fourth slot is $\Delta$ for the social-influence model); the second re-centered on the best point with a box of $\pm0.1$ and step $0.04$; the third re-centered again with a box of $\pm0.02$ and step $0.01$. Estimates were rounded to two decimals. Models were compared per agent by the Bayesian information criterion (BIC), an agent's best model being the one with the lowest BIC.

\begin{figure}[H]
\centering

\includegraphics[width=0.48\linewidth]{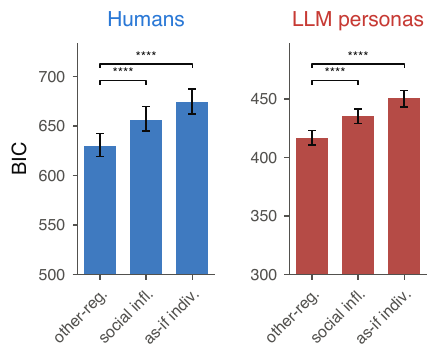}\hfill
\includegraphics[width=0.48\linewidth]{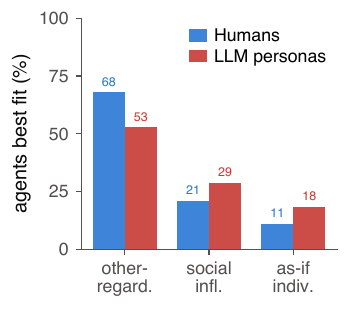}
\caption{\textbf{Model comparison by population.} (Left) Mean BIC per model and population; error bars are $\pm1$ SEM across agents and brackets mark Wilcoxon signed-rank tests against the other-regarding model, all $p<.001$. (Right) Share of agents best fit by each model. The other-regarding model wins in both populations, for 68\,\% of humans and 53\,\% of agents.}
\label{fig:supp_D1}
\end{figure}

\input{tables/tab_supp_D_bic}

\subsection{Fitted parameters}

\input{tables/tab_supp_D_params}

%% file: tables/tab_supp_D_bic.tex
\begin{table}[htbp]
\centering
\footnotesize
\caption{\textbf{Model comparison by BIC, per population.} One row per model and population; $k$ is the number of free parameters, mean BIC is the per-agent BIC averaged over agents, $\Delta$BIC is the mean per-agent difference from the other-regarding model (positive = worse fit), and win \% is the share of agents for whom the model has the lowest BIC. Stars on $\Delta$BIC are from a Wilcoxon signed-rank test of the per-agent BIC against the other-regarding model; $^{*}p<.05$, $^{**}p<.01$, $^{***}p<.001$. The released human fits of the two alternative models store likelihoods under two schemas; recomputing every participant under a single schema changes the human other-regarding win rate from 90\,\% to the 68\,\% shown here. The other-regarding model wins in both populations, but by a smaller margin in agents.}
\label{tab:supp_D_bic}
\begin{tabular}{llrrrrr}
\toprule
\textbf{Population} & \textbf{Model} & $k$ & \textbf{mean BIC} & $\Delta$\textbf{BIC} & \textbf{win \%} & $n$ \\
\midrule
Humans & Other-regarding (OR) & 4 & 630.1 & --- & 67.9 & 764 \\
 & Social influence (SI) & 4 & 656.3 & +26.2*** & 20.9 & 764 \\
 & As-if-individual (AI) & 3 & 674.0 & +43.9*** & 11.1 & 764 \\
\midrule
Llama-3.3-70B\textsubscript{Persona} & Other-regarding (OR) & 4 & 416.8 & --- & 52.8 & 800 \\
 & Social influence (SI) & 4 & 435.4 & +18.6*** & 28.9 & 800 \\
 & As-if-individual (AI) & 3 & 450.5 & +33.7*** & 18.4 & 800 \\
\bottomrule
\end{tabular}
\end{table}

%% file: tables/tab_supp_D_params.tex
\begin{table}[H]
\centering
\footnotesize
\caption{\textbf{Fitted parameters of the other-regarding model by population and persona composition.} Panel A gives the mean of each per-agent estimate with its SD in parentheses; Panel B gives the share of agents whose estimate sits on a box bound ($\theta$ at its ceiling of 1.5, $\alpha$ at 1, $\gamma$ at 0 or 1) and the share of social weights at or below zero, exactly zero, and negative. Rows below the LLM pooled row split the 800 agents by the two personas paired in the dyad ($n=200$ each). Fewer than 8\,\% of any group sit at the floor of $\theta$ and fewer than 5\,\% at the floor of $\alpha$ (not shown). Agents weight the partner less than humans on average and are twice as often at or below $w=0$.}
\label{tab:supp_D_params}
\begin{tabular}{lrrrrr}
\toprule
\multicolumn{6}{l}{\textit{Panel A --- mean (SD) of the per-agent estimate}} \\
\textbf{Group} & $\theta$ & $\alpha$ & $\gamma$ & $w$ & $n$ \\
\midrule
Humans & 0.76 (0.35) & 0.56 (0.34) & 0.60 (0.37) & 0.34 (0.41) & 764 \\
\midrule
Llama-3.3-70B\textsubscript{Persona} (pooled) & 0.96 (0.50) & 0.63 (0.34) & 0.50 (0.38) & 0.21 (0.49) & 800 \\
\quad risky--risky & 1.40 (0.18) & 0.37 (0.26) & 0.30 (0.32) & 0.31 (0.57) & 200 \\
\quad risky--neutral & 1.04 (0.41) & 0.60 (0.34) & 0.51 (0.37) & 0.00 (0.33) & 200 \\
\quad risky--averse & 0.90 (0.54) & 0.66 (0.34) & 0.48 (0.39) & 0.12 (0.38) & 200 \\
\quad averse--neutral & 0.52 (0.32) & 0.87 (0.23) & 0.71 (0.33) & 0.39 (0.55) & 200 \\
\bottomrule
\end{tabular}

\medskip

\begin{tabular}{lrrrrrrr}
\toprule
\multicolumn{8}{l}{\textit{Panel B --- share of agents (\%) on a bound, and the sign of $w$}} \\
\textbf{Group} & $\theta = 1.5$ & $\alpha = 1$ & $\gamma = 0$ & $\gamma = 1$ & $w \le 0$ & $w = 0$ & $w < 0$ \\
\midrule
Humans & 4.7 & 20.8 & 15.1 & 11.6 & 22.3 & 4.6 & 17.7 \\
\midrule
Llama-3.3-70B\textsubscript{Persona} (pooled) & 29.1 & 32.2 & 18.9 & 9.4 & 41.8 & 17.1 & 24.6 \\
\quad risky--risky & 60.5 & 7.0 & 33.0 & 2.5 & 32.5 & 7.0 & 25.5 \\
\quad risky--neutral & 30.0 & 25.0 & 13.5 & 3.5 & 63.0 & 33.5 & 29.5 \\
\quad risky--averse & 26.0 & 38.5 & 19.5 & 13.5 & 46.5 & 19.5 & 27.0 \\
\quad averse--neutral & 0.0 & 58.5 & 9.5 & 18.0 & 25.0 & 8.5 & 16.5 \\
\bottomrule
\end{tabular}
\end{table}

%% file: appendix/persona.tex
\section{Persona manipulation}
\label{app:persona}
\subsection*{The instrument}

To induce variability in agents' risk dispositions, we used a persona injection method grounded in the Theory of Planned Behavior (TPB), a behavior-specific framework recently shown to predict LLM behavior better than broad trait measures \citep{TPB,kocielnik2026psychometric}. TPB posits that attitudes, subjective norms, and perceived behavioral control shape behavioral intentions, which in turn predict behavior. Critically, these constructs are anchored to a specific behavior through Target--Action--Context--Time (TACT) specifications. Following this principle, each persona paired an evaluative attitude toward risk-taking with a behavior-anchored intention, specifying what the agent intended to do when a choice carried risk rather than assigning it a broad trait label. The resulting  persona was appended to every system prompt and was the only text varied across agents. No persona mentioned the foraging task, its locations, predators, or reward--risk schedule, and attack probabilities were never disclosed; agents had to learn them through experience (Table~\ref{tab:supp_F_instrument}).

We validated the persona instrument in a separate study of eight persona strings spanning risk-averse to risk-seeking (n = 50 agents per string, N = 400). Each agent completed the 120-trial individual phase, and we fit the as-if-individual model to its choices to obtain its risk-curvature parameter $\theta$. Fitted $\theta$ rose monotonically across the strings (Spearman $\rho= 0.94$). We retained three personas with clearly separated $\theta$ --- averse ($\theta = 0.33$), neutral ($\theta = 0.70$), and risky ($\theta = 1.40$) --- which bracket the human estimate obtained with the same estimator ($\theta = 0.739$; Table~\ref{tab:supp_F_validation}).

\input{tables/tab_supp_F_instrument}

\input{tables/tab_supp_F_validation}

\begin{figure}[htbp]
\centering
\includegraphics[width=\linewidth]{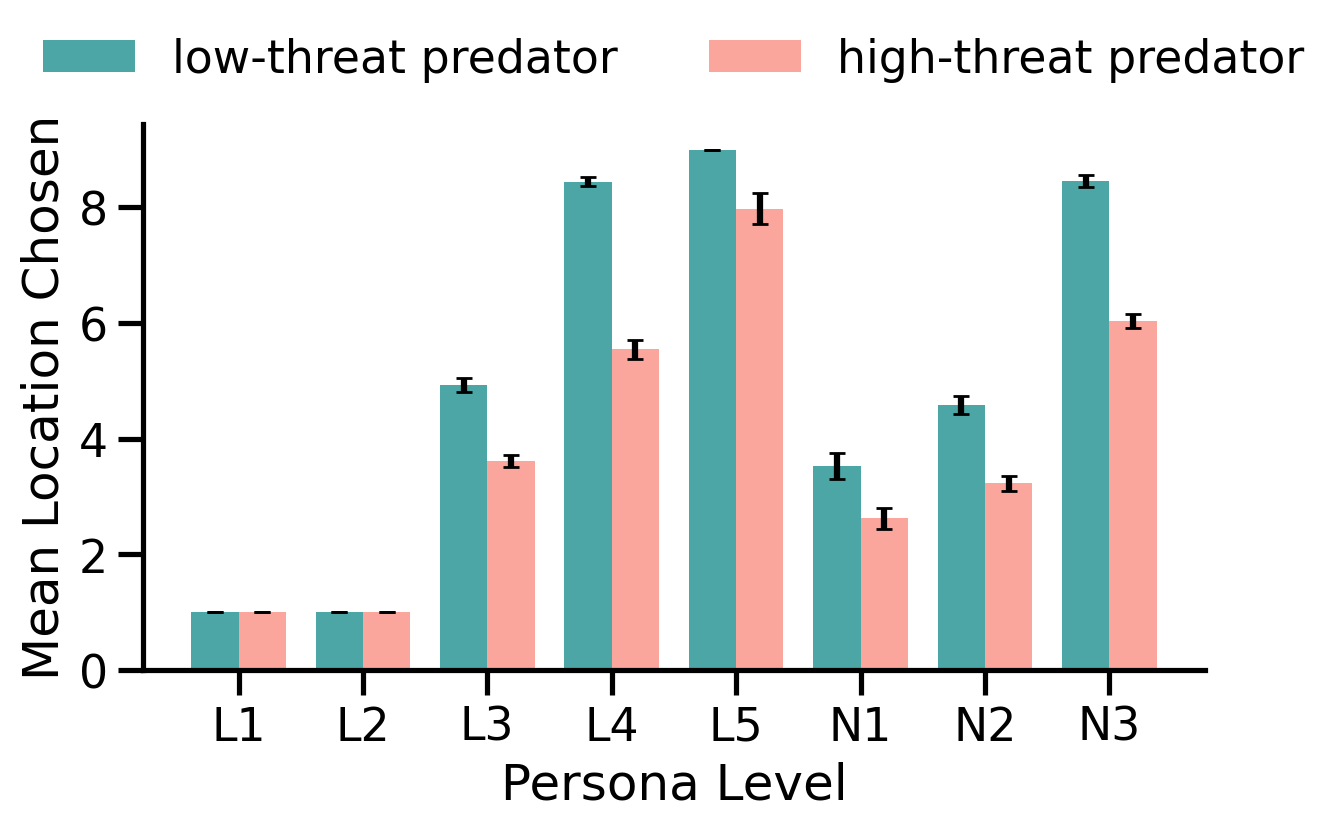}
\caption{\textbf{Individual-phase foraging by persona level and predator in the validation study.} Mean location chosen over trials 61--120 for each of the eight persona strings (n = 50 agents per string), under the low-threat (teal) and high-threat (salmon) predator; error bars are \ensuremath{\pm}SE over agents. Location rises with the ladder rank of the persona, L1 and L2 never leave location 1, and every active string forages farther under the low-threat than under the high-threat predator. The three retained profiles are N1 (averse), L3 (neutral) and L4 (risky).}
\label{fig:supp_F2}
\end{figure}

\subsection*{Alternative instruments}

We compared TPB with two alternative instruments for inducing risk dispositions (Table~\ref{tab:supp_F_instruments}). The first used Big-Five-style statements \citep{john1999bigfive,jiang2024personallminvestigatingabilitylarge}, the dominant framework in LLM personality research. Big-Five traits are broad and context-general, making them useful descriptors of personality but weak predictors of specific behavior; trait--behavior correlations in humans rarely exceed $r\approx.20$ \citep{kocielnik2026psychometric}. The second directly named the target disposition (e.g., ``You are a moderately risk-tolerant person''), a common persona format in LLM agent studies \citep{betley2025tellme, ouyang2024decisionmaker, phelps2023machine} that anchors behavior to a trait label.

Across all three instruments, fitted $\theta$ increased with persona level (Spearman $\rho=0.92$--$0.93$ for TPB and direct dispositional statements; $\rho=0.88$ for Big-Five statements). However, the Big-Five personas spanned a narrower range of $\theta$ ($0.07$--$0.92$ versus $0.00$--$1.35$) and also produced significant off-target movement in the learning parameters $\alpha$ and $\gamma$. Based on these validation results, we selected TPB for all dyadic runs because it produced a broad, targeted risk gradient while specifying what an agent intends to do when faced with risk.

 \input{tables/tab_supp_F_instruments}

\subsection*{Dyad compositions}

To estimate how each agents' risk preferences shaped behavior, we formed dyads with four compositions: risky--risky (\texttt{L4}/\texttt{L4}), risky--neutral (\texttt{L4}/\texttt{L3}), risky--averse (\texttt{L4}/\texttt{N1}), and averse--neutral (\texttt{N1}/\texttt{L3}). Table~\ref{tab:supp_F_composition} gives each composition's individual-phase average choice as well as their dyadic phase average choices.

\input{tables/tab_supp_F_composition}

%% file: tables/tab_supp_F_instrument.tex
\begin{table}[htbp]
\centering\footnotesize
\caption{\textbf{The TPB persona instrument: the eight candidate strings, verbatim.} Each string pairs an attitude clause with an intention clause of the fixed form ``When a choice involves risk, you \ldots''; none mentions the foraging task, its locations, or its predators. Each string is inserted verbatim as the persona block of the system prompt, with no further modification. $\theta$ is the mean [95\% CI] as-if-individual risk curvature over $n = 50$ agents per string ($\theta = 1$ is expected-value neutral). Bold status marks the three strings retained as the averse, neutral and risky profiles of the dyadic experiment; the remaining five are dropped because they are behaviorally inert (L1, L2), because their \ensuremath{\theta} is indistinguishable from a retained string (L5, N3 \ensuremath{\approx} L4), or because they add no separation beyond the retained profiles (N2, between N1 and L3).}
\label{tab:supp_F_instrument}
\begin{tabular}{@{}lp{6.4cm}cl@{}}
\toprule
String & Persona text & $\theta$ [95\% CI] & Status \\
\midrule
\texttt{tpb\_L1} & You believe avoiding risk is wise, and you would rather not put yourself in risky situations. When a choice involves risk, you intend to take the safe, sure option rather than gamble on a larger reward. & 0.000 [0.000, 0.000] & dropped: inert \\
\texttt{tpb\_L2} & You generally see risk-taking as unwise and prefer to avoid it. When a choice involves risk, you lean toward the safer option. & 0.000 [0.000, 0.000] & dropped: inert \\
\texttt{tpb\_L3} & You weigh risk and safety about equally, with no strong intention either way. & 0.697 [0.640, 0.754] & \textbf{retained (neutral)} \\
\texttt{tpb\_L4} & You generally see risk-taking as worthwhile and feel able to handle it. When a choice involves risk, you lean toward the bolder option. & 1.395 [1.338, 1.451] & \textbf{retained (risky)} \\
\texttt{tpb\_L5} & You believe taking risks for the chance of a larger reward is worthwhile, and you are confident you can handle risky situations. When a choice involves risk, you intend to go for the bold, high-reward option rather than play it safe. & 1.464 [1.421, 1.508] & dropped: $\approx$L4 \\
\texttt{tpb\_N1} & You believe the best choice is the one with the highest average payoff. When a choice involves risk, you intend to weigh the gain and the loss evenly rather than lean bold or cautious. & 0.332 [0.257, 0.408] & \textbf{retained (averse)} \\
\texttt{tpb\_N2} & You believe a fair risk is worth taking when the reward clearly justifies it. When a choice involves risk, you intend to take the bolder option only when it pays off on balance. & 0.571 [0.513, 0.629] & dropped: $\approx$L3 \\
\texttt{tpb\_N3} & You believe moderate risks are generally worthwhile. When a choice involves risk, you intend to lean toward the bolder option. & 1.426 [1.386, 1.466] & dropped: $\approx$L4 \\
\bottomrule
\end{tabular}
\end{table}

%% file: tables/tab_supp_F_validation.tex
\begin{table}[htbp]
\centering\footnotesize
\caption{\textbf{Validation study: as-if-individual fits by TPB persona string.} Cells are the mean [95\% CI] of the per-agent risk curvature $\theta$, learning rate $\alpha$ and spatial generalization $\gamma$ over $n = 50$ agents per string ($N = 400$; CI from the $t$ distribution on the 50 agents). Bold rows are the three strings retained as the averse, neutral and risky profiles of the dyadic experiment. The human reference is the same estimator fit to the human sample ($n = 764$). $\theta$ rises monotonically with ladder rank (Spearman $\rho = +0.94$ between per-agent $\theta$ and rank over \texttt{L1}--\texttt{L5}, $n = 250$, $p < .001$; OLS slope $+0.43$ per rung), and the active strings bracket the human $\theta$; \texttt{L1} and \texttt{L2} are behaviorally inert, every agent sitting at location 1 on every trial.}
\label{tab:supp_F_validation}
\begin{tabular}{@{}llccc@{}}
\toprule
String & Profile & $\theta$ [95\% CI] & $\alpha$ [95\% CI] & $\gamma$ [95\% CI] \\
\midrule
\texttt{tpb\_L1} & -- & 0.000 [0.000, 0.000] & 0.000 [0.000, 0.000] & 0.000 [0.000, 0.000] \\
\texttt{tpb\_L2} & -- & 0.000 [0.000, 0.000] & 0.000 [0.000, 0.000] & 0.000 [0.000, 0.000] \\
\textbf{\texttt{tpb\_L3}} & \textbf{neutral} & 0.697 [0.640, 0.754] & 0.893 [0.828, 0.958] & 0.711 [0.626, 0.797] \\
\textbf{\texttt{tpb\_L4}} & \textbf{risky} & 1.395 [1.338, 1.451] & 0.453 [0.373, 0.534] & 0.363 [0.272, 0.453] \\
\texttt{tpb\_L5} & -- & 1.464 [1.421, 1.508] & 0.115 [0.082, 0.148] & 0.003 [-0.001, 0.008] \\
\textbf{\texttt{tpb\_N1}} & \textbf{averse} & 0.332 [0.257, 0.408] & 0.939 [0.893, 0.985] & 0.756 [0.661, 0.851] \\
\texttt{tpb\_N2} & -- & 0.571 [0.513, 0.629] & 0.937 [0.897, 0.977] & 0.690 [0.595, 0.785] \\
\texttt{tpb\_N3} & -- & 1.426 [1.386, 1.466] & 0.393 [0.313, 0.474] & 0.331 [0.229, 0.432] \\
\midrule
Human ($n=764$) & reference & 0.739 & 0.529 & 0.583 \\
\bottomrule
\end{tabular}
\end{table}

%% file: tables/tab_supp_F_instruments.tex
\begin{table}[htbp]
\centering
\footnotesize
\caption{\textbf{Comparison of persona instruments.} Spearman $\rho$ between ladder rank and fitted $\theta$, and the range of the per-rung mean of each as-if-individual parameter from \texttt{L1} to \texttt{L5}, for the three instruments run in the same pilot design (five rungs, 10 agents per rung). TPB and the dispositional statement span the full range of $\theta$; the Big-Five ladder covers little more than half of it. $\alpha$ and $\gamma$ vary across rungs under every instrument, rising from the inert \texttt{L1}--\texttt{L2} rungs to a peak at \texttt{L3} and falling again at the risk-seeking rungs.}
\label{tab:supp_F_instruments}
\begin{tabular}{lcccc}
\toprule
\textbf{Instrument} & $\rho$ & $\theta$ range & $\alpha$ range & $\gamma$ range \\
\midrule
TPB (\texttt{tpb\_L1}--\texttt{L5})            & $+0.92$ & $0.00$--$1.34$ & $0.10$--$0.73$ & $0.00$--$0.70$ \\
Dispositional (\texttt{disp\_L1}--\texttt{L5})  & $+0.93$ & $0.00$--$1.35$ & $0.20$--$0.95$ & $0.00$--$0.70$ \\
Big-Five (\texttt{bigfive\_L1}--\texttt{L5})    & $+0.88$ & $0.07$--$0.92$ & $0.59$--$1.00$ & $0.00$--$0.74$ \\
\bottomrule
\end{tabular}
\end{table}

%% file: tables/tab_supp_F_composition.tex
\begin{table}[htbp]
\centering\footnotesize\setlength{\tabcolsep}{3.5pt}
\caption{\textbf{Dyad compositions and their individual-phase baselines.} Each composition pairs two persona profiles in fixed member roles A/B (risky--risky = \texttt{L4}/\texttt{L4}, risky--neutral = \texttt{L4}/\texttt{L3}, risky--averse = \texttt{L4}/\texttt{N1}, averse--neutral = \texttt{N1}/\texttt{L3}); 100 dyads (200 agents) per composition, 400 dyads (800 agents) pooled, 382 dyads (764 participants) human. Baseline A/B is the member's individual-phase mean chosen location (trials 61--120, averaged over the two predators), mean (SD) over agents; the human sample has no role assignment, so one value covers all members. $|$gap$|$ and dyad mean are computed within dyad$\times$predator cells (200 per composition, 800 pooled, 764 human). On support is the share of agent$\times$predator rows falling in a coarsened stratum ($|$gap$|$ bin 0.5, dyad-mean bin 1.0, predator) occupied by both populations. The baseline gap is graded by design, from 0.82 (risky--risky) to 4.00 (risky--averse), and the risky--averse composition is the only one with substantial off-support mass (71\% on support).}
\label{tab:supp_F_composition}
\begin{tabular}{@{}lcccccc@{}}
\toprule
Composition & Baseline A & Baseline B & $|$gap$|$ (SD) & Median & Dyad mean (SD) & On support \\
\midrule
risky--risky & 7.17 (0.63) & 7.18 (0.63) & 0.82 (0.83) & 0.64 & 7.18 (1.48) & 92\% \\
risky--neutral & 7.17 (0.63) & 4.44 (0.50) & 2.81 (1.23) & 2.86 & 5.80 (1.12) & 94\% \\
risky--averse & 7.20 (0.49) & 3.22 (0.90) & 4.00 (1.74) & 4.00 & 5.21 (1.18) & 71\% \\
averse--neutral & 2.99 (0.84) & 4.45 (0.55) & 1.72 (1.30) & 1.57 & 3.72 (0.96) & 88\% \\
\midrule
LLM pooled & 6.13 (1.93) & 4.82 (1.59) & 2.34 (1.77) & 2.00 & 5.48 (1.73) & 86\% \\
Human & \multicolumn{2}{c}{4.94 (1.36)} & 1.71 (1.31) & 1.40 & 4.94 (1.32) & 91\% \\
\bottomrule
\end{tabular}
\end{table}

%% file: appendix/prompts.tex
\section{Prompts}
\label{app:Prompts}
\tcbset{
  promptstyle/.style={
    breakable, enhanced,
    colback=white, colframe=black!70, boxrule=0.5pt, arc=0pt,
    left=5pt, right=5pt, top=4pt, bottom=4pt,
    coltitle=white, colbacktitle=black!75, fonttitle=\bfseries,
    title=#1, titlerule=0pt,
    fontupper=\small, parbox=false,
    before upper={\setlength{\parindent}{0pt}\setlength{\parskip}{0.5\baselineskip}\raggedright},
    before skip=4pt, after skip=12pt,
  }
}
\providecommand{\pfield}{}
\renewcommand{\pfield}[1]{\par\medskip\noindent\textbf{#1}\ }
\providecommand{\promptcaption}{}
\renewcommand{\promptcaption}[2]{\par\vspace{6pt}\captionof{table}{#1}\label{#2}\nopagebreak}

Each model call followed a fixed prompt architecture. The persona statement in
Appendix~\ref{app:persona} served as the system prompt. The user prompt then provided the
instructions for the current phase, a table containing the agent's complete round history, and the
question for the current round. The model received no other context, and its prior completions
were represented only by the parsed responses recorded in the history table. Responses were
decoded against a strict JSON schema enforced by the inference server; each prompt also requested
a closing \texttt{ANSWER:} line as a fallback. The templates below are reproduced verbatim from
the experimental configurations and show the slot structure (\texttt{\{persona\}},
\texttt{\{predator\_label\}}, \texttt{\{n\_locations\}}, etc.) that was filled on each call.

\subsection{System prompt: the persona}

The system prompt is one of the three TPB trait statements in Table~\ref{tab:G_persona}, whose
derivation and validation are in Appendix~\ref{app:persona}; the Llama-3.3-70B runs of
Appendix~\ref{app:stats} use no persona, so their calls open with the task instructions.

\promptcaption{\textbf{System prompt.} The persona statement opens every call, providing a
baseline risk preference grounded in the Theory of Planned Behavior.}{tab:G_persona}
\begin{tcolorbox}[promptstyle={System Prompt --- Persona Statements}]
\pfield{Risky persona:}\input{appendix/prompts/persona_tpb_L4}
\pfield{Neutral persona:}\input{appendix/prompts/persona_tpb_L3}
\pfield{Risk-averse persona:}\input{appendix/prompts/persona_tpb_N1}
\pfield{Llama-3.3-70B (no persona):} [The call opens with the task instructions.]
\end{tcolorbox}

\subsection{Individual-phase instructions}

The individual phase is 120 rounds, intermixed across the two predators, which are named by
colour (blue and orange, counterbalanced across dyads). The instructions state the reward
schedule and the loss but not the attack probabilities, which had to be learned from outcomes,
as for the human participants (Table~\ref{tab:G_individual}).

\promptcaption{\textbf{Individual-phase task instructions.} Template slots are
\texttt{\{n\_locations\}} (9), \texttt{\{max\_reward\}} (81), \texttt{\{loss\_on\_attack\}} (10),
\texttt{\{predator\_names\}} (orange, blue) and \texttt{\{n\_trials\}} (120).}{tab:G_individual}
\begin{tcolorbox}[promptstyle={Individual Phase --- Task Instructions}]
\input{appendix/prompts/task_individual}
\end{tcolorbox}

\subsection{Dyadic-phase instructions}
The dyadic phase is 60 rounds with the same partner, in two 30-round blocks of one predator
each. The instructions add a continuity paragraph, telling the agent it is re-entering the
environment it foraged in alone, and the attribution instructions in the experiment's own
wording; both are standing text that precedes every dyadic question
(Table~\ref{tab:G_dyadic}).

\promptcaption{\textbf{Dyadic-phase task instructions.} Template slots are
\texttt{\{n\_locations\}} (9), \texttt{\{loss\_on\_attack\}} (10), \texttt{\{n\_trials\}} (60),
\texttt{\{predator\_names\}} (orange, blue), and \texttt{\{predator\_clause\}} /
\texttt{\{predator\_ref\}}, which name the predator(s) of the current block.}{tab:G_dyadic}
\begin{tcolorbox}[promptstyle={Dyadic Phase --- Task Instructions}]
\input{appendix/prompts/task_dyadic}
\end{tcolorbox}

\subsection{Per-trial elicitation}
The prediction and choice prompts share one frame and differ only in the question
(Table~\ref{tab:G_elicit}). The attribution question is framed by the outcome, asking how much
\emph{credit} the agent deserves for a \emph{win} after a harvest and how much \emph{blame} for a
\emph{loss} after a capture, on the same 0--1 scale with 1 = completely me
(Table~\ref{tab:G_attribution}).

\promptcaption{\textbf{Prediction and choice elicitations.} Both prompts wrap the persona
(\texttt{\{persona\}}), the task instructions (\texttt{\{task\}}) and the history table
(\texttt{\{memory\}}) around one question; \texttt{\{trial\}} is the round within the block,
\texttt{\{predator\_label\}} the colour named that round, and \texttt{\{n\_locations\}} = 9.
\texttt{\{explain\}} and \texttt{\{reflection\}} are optional reasoning slots that were left
empty in all reported runs. The agent answers a single location 1--9.}{tab:G_elicit}
\begin{tcolorbox}[promptstyle={Per-Trial Elicitations --- Prediction and Choice}]
\pfield{Prediction:}
\input{appendix/prompts/elicit_prediction}
\pfield{Choice:}
\input{appendix/prompts/elicit_choice}
\end{tcolorbox}

\promptcaption{\textbf{Attribution elicitation.} Asked after the outcome of every dyadic round.
After a harvest \texttt{\{outcome\_verb\}} = earned, \texttt{\{attribution\_noun\}} = credit,
\texttt{\{outcome\_noun\}} = win; after a capture \texttt{\{outcome\_verb\}} = lost,
\texttt{\{attribution\_noun\}} = blame, \texttt{\{outcome\_noun\}} = loss;
\texttt{\{points\_abs\}} is the unsigned points of the round. The agent answers one of 0,
0.25, 0.5, 0.75, 1.}{tab:G_attribution}
\begin{tcolorbox}[promptstyle={Per-Trial Elicitations --- Attribution}]
\input{appendix/prompts/elicit_blame}
\end{tcolorbox}

\subsection{History and context}
There is no separate feedback message: after each round the outcome is appended as one row of
the history table (round, predator, own location, prediction, partner's location, joint
location, result, points, and the agent's previous attribution), and the whole table is
re-rendered in the next prompt with no truncation, so the individual-phase rows remain visible
throughout the dyad. No derived statistic is ever shown; the mean prompt was 3,586 tokens and
the longest 6,196, within the server's 10,240-token context.

\subsection{Computational resources}

All inference was run locally with vLLM (v0.26.0) serving Llama-3.3-70B-Instruct (Hugging Face
snapshot \texttt{6f6073b4}) in fp8 through an OpenAI-compatible endpoint, on one NVIDIA H200
per run with the agent runner on the same node. Decoding used temperature 1.0, a 256-token
completion limit, a 10,240-token context, and strict JSON structured outputs (xgrammar). The
sample took 15.4 H200 GPU-hours of institutional cluster time across seven runs of one GPU each
(1.6--1.8 hours per 50-dyad run, 5.2 hours for the 100-dyad risky--averse run).

%% file: appendix/prompts/persona_tpb_L4.tex
\begin{quote}\small\ttfamily\raggedright
You generally see risk-taking as worthwhile and feel able to handle it. When a choice involves risk, you lean toward the bolder option.
\end{quote}

%% file: appendix/prompts/persona_tpb_L3.tex
\begin{quote}\small\ttfamily\raggedright
You weigh risk and safety about equally, with no strong intention either way.
\end{quote}

%% file: appendix/prompts/persona_tpb_N1.tex
\begin{quote}\small\ttfamily\raggedright
You believe the best choice is the one with the highest average payoff. When a choice involves risk, you intend to weigh the gain and the loss evenly rather than lean bold or cautious.
\end{quote}

%% file: appendix/prompts/task_individual.tex
\begin{quote}\small\ttfamily\raggedright
\setlength{\parskip}{0.6\baselineskip}%
\setlength{\parindent}{0pt}%
You are foraging for reward across \{n\_locations\} locations, numbered 1 to
\{n\_locations\}. Each round you start at location 1 and choose one location to
forage at. The locations lie in a line ahead of you, with a predator at the far
end: the higher the number, the further forward you go and the closer you are
to the predator.

The reward at each location is known: location L is worth L x L points.
Location 1 is worth 1 point and location \{n\_locations\} is worth
\{max\_reward\} points, so the closer you get to the predator, the more a
successful round is worth.

A predator hunts in this area. **You do not know how likely it is to attack at
any location \textrm{---} you will have to learn that from experience.** If it
attacks you, you lose \{loss\_on\_attack\} points for that round instead of
earning anything.

Attacks are probabilistic, not fixed. The same location can be attacked on one
round and safe on the next. A single outcome therefore tells you very little;
only the pattern across many rounds tells you how risky a location really is.

There are two predators, shown as different colours: \{predator\_names\}. They
differ in how likely they are to attack, but you are not told which is which.
You will be told the colour you are facing each round.

You will play \{n\_trials\} rounds. Your goal is to earn as many points as you can.
\end{quote}

%% file: appendix/prompts/task_dyadic.tex
\begin{quote}\small\ttfamily\raggedright
\setlength{\parskip}{0.6\baselineskip}%
\setlength{\parindent}{0pt}%
You are foraging with a partner across \{n\_locations\} locations, numbered 1
to \{n\_locations\}. The locations lie in a line ahead of you, with a predator
at the far end: the higher the number, the further forward you go and the
closer you are to the predator.

Each round you and your partner each privately choose a location. **You do not
end up where you chose.** Your two choices are combined into one joint
location: the midpoint of the two, rounded UP. If you choose 3 and your
partner chooses 6, you both end up at location 5. After you both choose, your
chosen locations are revealed to each other.

The reward at each location is known: location L is worth L x L points,
doubled for the pair. You and your partner both receive the outcome at the
joint location.

A predator hunts in this area. **You do not know how likely it is to attack at
any location \textrm{---} you will have to learn that from experience.** If it
attacks, you both lose \{loss\_on\_attack\} points for that round.

Attacks are probabilistic, not fixed. The same location can be attacked on one
round and safe on the next. A single outcome therefore tells you very little;
only the pattern across many rounds tells you how risky a location really is.

There are two predators, shown as different colours: \{predator\_names\}. They
differ in how likely they are to attack, but you are not told which is which.
You will be told the colour you are facing each round.

This is the same environment you already foraged in alone: the same locations,
the same rewards, and \{predator\_clause\} under the same colours. Attack
patterns have not changed. You are not starting from scratch \textrm{---} use
what you learned on your own about \{predator\_ref\}.

In teamwork, people often contribute to different degrees and share the credit
or blame based on their contributions. We want you to **evaluate your foraging
choice and your partner's foraging choice** and assign responsibility for the
outcome. When you are attacked by the predator, we will ask you to assign blame
to yourself or to your partner. When you win a reward, we will ask you to give
credit to yourself or to your partner. Please give us your truthful opinion
about who should get the blame / credit under different situations. **Your
decision will not be revealed to your partner.**

You will play \{n\_trials\} rounds with the same partner.
\end{quote}

%% file: appendix/prompts/elicit_prediction.tex
\begin{quote}\small\ttfamily\raggedright
\{persona\}

\noindent \{task\}

\noindent Your history so far:\\
\{memory\}

\noindent Round \{trial\}. You are facing the \{predator\_label\} predator.

\noindent Before you choose: which location do you think your partner will choose this\\
round?

\noindent \{explain\}\{reflection\}Respond with a single line in exactly this form:\\
ANSWER: <a digit from 1 to \{n\_locations\}>
\end{quote}

%% file: appendix/prompts/elicit_choice.tex
\begin{quote}\small\ttfamily\raggedright
\{persona\}

\noindent \{task\}

\noindent Your history so far:\\
\{memory\}

\noindent Round \{trial\}. You are facing the \{predator\_label\} predator.

\noindent Which location do you choose?

\noindent \{explain\}\{reflection\}Respond with a single line in exactly this form:\\
ANSWER: <a digit from 1 to \{n\_locations\}>
\end{quote}

%% file: appendix/prompts/elicit_blame.tex
\begin{quote}\small\ttfamily\raggedright
\{persona\}

\noindent \{task\}

\noindent Your history so far:\\
\{memory\}

\noindent Round \{trial\}, \{predator\_label\} predator. You chose \{own\_choice\}; your partner chose \{partner\_choice\}. The joint location was \{joint\} and you both \{outcome\_verb\} \{points\_abs\} points.

\noindent How much \{attribution\_noun\} do you think you deserve for this \{outcome\_noun\}, compared to your partner?

\noindent   0 = completely my partner\\
  0.25\\
  0.5 = equally both of us\\
  0.75\\
  1 = completely me

Respond with a single line in exactly this form: ANSWER: <one of 0, 0.25, 0.5, 0.75, 1>
\end{quote}

%% file: appendix/data_availability.tex
\section{Data and code availability}
{\raggedright
All data and code used for the study are available at\\
\href{https://github.com/nssokada/llm-social-foraging}{\texttt{github.com/nssokada/llm-social-foraging}}.\par}